\def\eq#1{Eq.\,(\ref{#1})}
\def\nl{\hfil\break\noindent}

\documentclass[amsmath,amssymb,superscriptaddress,showkeys,showpacs,nofootinbib]{revtex4}
\usepackage{amsmath}
\usepackage{bm}
\usepackage{graphicx}
\usepackage{rotating}
\usepackage{mathrsfs} %Script characters: {\mathscr D}

\begin{document}
% ---------------------------
%\hspace*{5 in}CUQM - 139
%\vskip 0.4 in
%\hfill fick [5$^{\rm th}$ May 2012]\vspace*{0.4in}
% ---------------------------
%\def\ttitle{Derivation of a potential model for the Minkowski-space Bethe--Salpeter equation}
%\def\ttitle{Potential-model approach to solutions of the Minkowski-space Bethe--Salpeter equation}
%\def\ttitle{Schr\"odinger Model for Solutions of the Minkowski-Space Bethe--Salpeter Equation}
\def\ttitle{Schr\"odinger models for solutions of the Bethe--Salpeter equation in Minkowski space}
\title{\ttitle}
\markboth{R.~L.~Hall \& W.~Lucha}{\ttitle}

\author{Richard L.~Hall}
\email{rhall@mathstat.concordia.ca}
\affiliation{Department of Mathematics and Statistics, Concordia University,
1455 de Maisonneuve Boulevard West, Montr\'eal,
Qu\'ebec, Canada H3G 1M8}
\author{Wolfgang~Lucha}
\email{wolfgang.lucha@oeaw.ac.at}\affiliation{Institute for High
Energy Physics, Austrian Academy of Sciences, Nikolsdorfergasse
18,\\A-1050 Vienna, Austria}

\begin{abstract}\noindent By application of the `geometric spectral
inversion' technique, which we have recently generalized to
accommodate also {\em singular\/} interaction potentials, we
construct from spectral data emerging~from the solution of the
Minkowski-space formulation of the homogeneous Bethe--Salpeter
equation describing bound states of two spinless particles a
Schr\"odinger approach to such states in terms of nonrelativistic
potential models. This spectrally equivalent modeling of bound
states yields their {\em qualitative\/} features (masses, form
factors, etc.) without having to deal with the more involved
Bethe--Salpeter formalism.\end{abstract}

\keywords{Bethe--Salpeter equation, nonrelativistic potential
model, geometric spectral inversion}

\pacs{11.10.St, 03.65.Pm, 03.65.Ge.}

\maketitle

%%%%%%%%%%%%%%%%%%%%%%%%%%%%%%%%%%%%%%%%%%%%%%%%%%%%%%%%
\section{Introduction: Motivation and Incentive}
%%%%%%%%%%%%%%%%%%%%%%%%%%%%%%%%%%%%%%%%%%%%%%%%%%%%%%%%
Within the framework of relativistic quantum field theory, the
appropriate tool for the description of bound states~is, in
principle, the Bethe--Salpeter (BS) formalism
\cite{BSE_APS,GML,BSE}. In this approach, a bound state ${\rm
B}(P),$ of momentum $P$~and~mass $M,$ is described by its BS
amplitude, which in configuration-space representation is defined
by the matrix element of the time-ordered product of the field
operators of all bound-state constituents between vacuum
$|0\rangle$ and bound~state~$|{\rm B}(P)\rangle.$ In
momentum-space representation, the BS amplitude, upon splitting
off the center-of-momentum motion of the bound state and
suppression of all indices generically denoted by $\Phi(p,P),$
encodes the distribution of the relative momenta $p$ of the
bound-state constituents. It satisfies a formally exact BS
equation involving two kinds of dynamical~ingredients, namely, for
bound states composed of $n$ constituents, (a) the propagators
$S_i(p_i)$ ($i=1,2,\dots,n$) of the $n$ constituents of respective
individual momenta $p_i$ and (b) its BS interaction kernel $K,$ a
fully truncated $(2\,n)$-point Green function of the $n$
bound-state constituents, perturbatively defined as the sum (of
the countable infinity) of all `BS-irreducible'~Feynman graphs for
$n$-particle into $n$-particle scattering. For two bound-state
constituents, the BS equation is of the generic~form
\begin{equation}\Phi(p,P)=\frac{{\rm i}}{(2\pi)^4}\,S_1(p_1)\int{\rm
d}^4q\,K(p,q,P)\,\Phi(q,P)\,S_2(-p_2)\
.\label{Eq:BSE}\end{equation}

Physical considerations provide a profound motivation to formulate
one's BS framework in {\em Minkowski space\/}, with~the
pseudo-Euclidean space-time metric tensor $g_{\mu\nu}={\rm
diag}(+1,-1,-1,-1)$. In the Minkowski-space formulation, however,
finding solutions to the BS equation may be heavily impeded by the
presence of singularities induced by the propagators of the
bound-state constituents or its BS interaction kernel. As a
remedy, by assuming that analytic continuation of the
Minkowski-space formalism is possible and the Cauchy integral
theorem is applicable, it has been proposed~to~study the BS
equation in Euclidean space, with metric
$g_{\mu\nu}=\delta_{\mu\nu},$ reached by a procedure misleadingly
labeled Wick~`rotation' \cite{Wick}. The Euclidean-space
formulation facilitates making contact with lattice field theory,
usually defined in Euclidean space. Solutions of the BS equation
provide the set of mass eigenvalues $M$ and associated BS
amplitudes $\Phi$ of~the~bound~states. The mass eigenvalues
arising in Minkowski-space and Euclidean-space formulations of a
given BS equation are identical. For BS amplitudes, however,
complicated analyticity structures of the BS equation in the
complex plane cause~troubles: The BS solutions derived in one
formulation may differ from those obtained in the other one. That
is to say,~the~analytic continuation to Minkowski space of some
solution to the BS equation in Euclidean space may bear no
resemblance to its counterpart found as solution to the same BS
equation in Minkowski space. Since the BS amplitudes
determine~physical observables such as decay constants and form
factors, knowledge of the Minkowski-space amplitudes is highly
desirable.

This dilemma between, on the one hand, the comparative ease of
deriving solutions to a Euclidean-space BS~equation and, on the
other hand, the need of physical applications for BS amplitudes
constituting solutions to a Minkowski-space BS equation can be
tentatively resolved by developing---of course, only approximately
equivalent---Schr\"odinger~models. This may be effected by fixing
the interaction potential entering in the Schr\"odinger
Hamiltonian by spectral inversion of the bound-state mass
eigenvalues $M$ arising from the easier-to-accomplish solution of
the Euclidean-space BS equation. The wave functions obtained as
solutions of the resulting Schr\"odinger equation allow us to
compute decay~constants and form factors of bound states, provided
we succeed in acquiring control of the uncertainties introduced by
such~modeling. Obviously, we may estimate the accuracy of the
envisaged Schr\"odinger models by applying our proposal for
resolution of the dilemma to some known solutions of the
Minkowski-space BS equation in order to extract an
associated~Schr\"odinger potential and comparing the outcome of
the corresponding Schr\"odinger equation with the findings of the
BS framework.

Recently, renewed attempts of solving the BS equation for
two-particle bound states in Minkowski-space formulation have been
undertaken
\cite{CK05a,CK05b,CK05c,CK06,CK07,CK08,CK10a,CK10b,CK10c}. The
basic idea advocated for in Ref.~\cite{CK05a} is to remove the
singularities of the BS~amplitudes by considering an equivalent
integral transform of the BS equation (\ref{Eq:BSE}) obtained by
projection onto the light-front plane, and to take advantage of a
particular integral representation of the BS amplitude $\Phi(p,P)$
proposed by Nakanishi~\cite{Nakanishi}, in order to obtain a
nonsingular integral equation that is straightforward to solve
numerically. This route has~been applied to bound states
consisting either of two identical spin-$0$ bosons
\cite{CK05a,CK05b,CK05c,CK06,CK07,CK08} or of a spin-$\frac{1}{2}$
fermion and its antiparticle~\cite{CK10a,CK10b,CK10c}. Any such
bound state arises, in the case of two scalar constituents, from
their couplings to a further scalar boson or, in the case of
fermionic constituents, from their couplings to a scalar, a
pseudoscalar, or a massless vector~boson;~for~bound states of
scalars, the interactions are taken into account in the BS
equation by considering the BS kernel either in ladder
approximation, which amounts to the iteration of single-boson
exchanges, or in ladder-plus-cross-ladder approximation. Moreover,
for the bound-state constituents the studies
\cite{CK05a,CK05b,CK05c,CK06,CK07,CK08,CK10a,CK10b,CK10c} employ,
for simplicity, the free-propagator approximation.

In order to get an idea of the behaviour of such Schr\"odinger
interaction potentials to be expected to arise in the course of
spectral inversion, we recall that there is a well-paved path of
simplifications leading from the relativistic~BS~equation to its
nonrelativistic or `static' Schr\"odinger reduction. The sequence
of necessary steps involves several well-defined and thoroughly
studied approximations to the BS formalism (for brief reviews of
this reduction, consult, e.g., Refs.~\cite{Lucha91,
Lucha:Oberwoelz,Lucha:Dubrovnik}):\begin{enumerate}\item In some
instantaneous limit, realizable if in the bound-state's
center-of-momentum frame fixed by $P=(M,\mathbf{0})$ the BS kernel
takes the form $K(p,q,P)=K(\bm{p},\bm{q}),$ the BS equation may be
reduced to the {\em instantaneous~BS~equation\/} (for attempts in
these directions, consult, for instance,
Ref.~\cite{Lucha05:IBSEWEP} and references therein) for the
Salpeter amplitude$$\phi(\bm{p})\equiv\frac{1}{2\pi}\int{\rm
d}p_0\,\Phi(p)\ .$$\item The additional assumption of free
propagation of all bound-state constituents with effective masses
encompassing the dynamical self-energy effects leads to the {\em
Salpeter equation\/} \cite{SE}. (Note, however, that in quantum
field theory the Dyson--Schwinger equations relate every $n$-point
Green function to at least one $(m>n)$-point~Green~function. This
means, in particular, that the propagators, i.e., the 2-point
Green functions, and the $n$-point Green functions entering in the
BS kernel cannot be chosen independently: the use of free
propagators might be incompatible with the feature of confinement
exhibited by quantum chromodynamics, the theory describing the
strong interactions.)\item Dropping all negative-energy
contributions simplifies Salpeter's equation to the {\em reduced
Salpeter equation\/}
\mbox{\cite{Henriques76,Jacobs87,Gara89,Gara90,Lucha92C}}.\item
Furthermore, ignoring all spin degrees of freedom of all
bound-state constituents and assuming the BS interaction kernel
$K$ to be of convolution type, i.e., to depend only on the
difference of the involved relative momenta $\bm{p}$ and $\bm{q},$
$K(\bm{p},\bm{q}) =K(\bm{p}-\bm{q}),$ yields the {\em spinless
Salpeter equation}. Therein, the interactions manifest in form of
a potential arising, in configuration space, as the Fourier
transform of this kernel $K(\bm{p}-\bm{q}).$ This bound-state
equation~may~be viewed as a generalization of the Schr\"odinger
equation towards relativistic kinematics. Concise reviews of
various aspects and facets of semirelativistic approaches to the
bound-state problem may~be~found~in,~e.g.,
Refs.~\cite{Lucha92,Lucha94:Como,Lucha04:TWR}.\footnote{A related
approach is the quasipotential formalism devised by Todorov
\cite{Todorov}.}\item In an ultimate static limit, replacing in
the latter equation of motion the relativistic form of all
one-particle kinetic energies by the corresponding nonrelativistic
approximation, we eventually end up with the {\em
Schr\"odinger~equation}.\end{enumerate}

Let us begin our analysis by inspecting the simplest case: bound
states of two scalar constituents
\cite{CK05a,CK05b,CK05c,CK06,CK07,CK08}. In the ladder
approximation, the only contribution to the BS interaction kernel
derives from single-particle exchange.~Apart from the couplings of
the exchanged particle to the bound-state constituents, the BS
kernel is then nothing but the propagator of the exchanged
particle. For a scalar boson with mass $\mu,$ its free propagator
is given by $S(k)={\rm i}\,(k^2-\mu^2)^{-1}.$~~The~Fourier
transform in three dimensions of the instantaneous approximation
to this propagator, i.e., of its remnant ${\rm
i}\,(\bm{k}^2+\mu^2)^{-1},$ is proportional to the
configuration-space Yukawa potential $V(r)=-\exp(-\mu r)/r,$ thus
singular at the origin $r\equiv|\bm{x}|=0.$ As a consequence,
depending, clearly, on the proximity of the system described by
the BS equation to the nonrelativistic Schr\"odinger limit, the
outcome of any spectral inversion may be potentials resembling, to
some extent, the Yukawa type but modified, of course, by the
various effects ignored on the way down to the static limit, such
as relativistic~kinematics

\newpage\noindent or higher-order contributions to the BS
interaction kernel; cross-ladder terms are but the simplest
example of the latter. Accordingly, we have to devise and utilize
an inversion technique that is capable of dealing also with
singular~potentials.

To this end, we recently generalized the earlier geometric
spectral inversion \cite{inv1, inv2, inv3, inv4, inv5} to treat
singular potentials~\cite{FIS}. We suppose that $f(r)$ is the
shape of the potential and $v>0$ is the coupling parameter in the
Schr\"odinger Hamiltonian $H =-\Delta/(2m)+v\,f(r).$ In this
inversion technique, a functional sequence is built which starts
from a seed $f^{[0]}(r)$ and reconstructs the potential shape
$f(r)$ from a given spectral function $E = F(v)$ that defines how
a discrete eigenvalue~$E$ of $H$ depends on the parameter $v.$ The
most relevant earlier paper is Ref.~\cite{FIS}, which also
includes a proof of uniqueness for the inverse for a large class
of singular potentials. Here, the inversion sequence is defined in
Sect.~\ref{invseq}; a statement of the uniqueness theorem may be
found in Sect.~\ref{unique}. The principal goal of the present
paper is to take for $F(v)$ the solutions to the Minkowski-space
BS equation \cite{CK05a,CK05b,CK05c,CK06,CK07,CK08,CK10c} for
bound states of two scalar constituents, and to reconstruct
directly from this set of data an effective potential shape $f(r)$
in the Schr\"odinger model defined by the Hamiltonian~$H.$

The outline of this paper is as follows. In Sect.~\ref{Sec:GSI},
we summarize enough of the geometric spectral inversion theory~to
make this paper essentially self-contained. In
Sect.~\ref{Sec:CKdata}, we present the spectral data $F(v)$ from
the BS solutions \cite{CK05a,CK05b,CK05c,CK06,CK07,CK08}. In
Sect.~\ref{Sec:EP}, we apply the functional inversion sequence of
Sect.~\ref{invseq} to these data to construct the effective
potential shape. In Appendix~\ref{ANRR}, we sketch the
nonrelativistic reduction of the Bethe--Salpeter formalism for
scalar bound-state constituents along a route which mimics to the
utmost possible extent the case of fermionic bound-state
constituents.

%%%%%%%%%%%%%%%%%%%%%%%%%%%%%%%%%%%%%%%%%%%%%%%%%%%%%%%%%%%%%%%
\section{Geometric spectral inversion}\label{Sec:GSI}
%%%%%%%%%%%%%%%%%%%%%%%%%%%%%%%%%%%%%%%%%%%%%%%%%%%%%%%%%%%%%%%
We consider the discrete spectrum of a Schr\"odinger Hamiltonian operator
\begin{equation}\label{hamiltonian}
H = -\Delta +v\,f(r)\ ,\qquad r\equiv \|\bm{x}\|\ ,
\end{equation}
where $f(r)$ is the shape of an attractive central potential, and
$v >0$ is a coupling parameter. We shall assume~that~the
potentials are monotone non-decreasing and no more singular than
the Coulomb potential $f(r) = -1/r.$ The arguments we use apply
generally to the problem in $d>1$ spatial dimensions, but, for
definiteness, we shall usually assume that $d = 3.$ The operator
inequality \cite{GS,RS2}
\begin{equation}\label{opineq}
-\Delta \ge \left(\frac{d/2-1}{r}\right)^2\ ,\qquad d\ge3\ ,
\end{equation}
implies that a discrete spectrum exists for sufficiently large
coupling $v >0$. For $d=3,$ the Hamiltonian $H$ is bounded below
by
\begin{equation}\label{elower}
E \ge \min_{r > 0}\left[\frac{1}{4\,r^2} + v\,f(r)\right],
\end{equation}
and a simple trial function can be used to establish an upper
bound to $E.$ Thus, we may assume, in particular,~that~the
ground-state energy may be written as a function $E = F(v)$. An
explicit example of the class of problems we consider~is provided
by the Hulth\'en potential, whose shape is given by $f(r) =
-1/(e^{r}-1)$ and whose s-state ($\ell = 0$) eigenvalues~$E_n$ are
given \cite{flug} exactly for $d=3$ by the formula
\begin{equation}\label{ehulthen}
E_n = F_n(v) = -\left(\frac{v-n^2}{2\,n}\right)^2\ ,\qquad v >
n^2\ ,\qquad n = 1,2,3,\dots\ .
\end{equation}
The problem discussed in the present paper may be stated as
follows: given, for example, the curve $F_1(v),$ can we use this
spectral data to reconstruct the potential shape $f(r)$? We call
this reconstruction a `geometric spectral inversion'.

%%%%%%%%%%%%%%%%%%%%%%%%%%%%%%%%%%%%%%%%%%%%%%%%%%%%%%%%%%%%%%%%%%%%%%%%%%%%%%
\subsection{Exact representation of spectral functions by kinetic potentials}
%%%%%%%%%%%%%%%%%%%%%%%%%%%%%%%%%%%%%%%%%%%%%%%%%%%%%%%%%%%%%%%%%%%%%%%%%%%%%%
The discrete spectra of operators bounded from below,
such as $H = -\Delta + v\,f(r),$ may be characterized
variationally \cite{RS4}. Thus, the ground-state energy may be
written
\begin{equation}\label{varchar}
F(v) = \inf_{{{\scriptstyle \psi \in {\cal D}(H)} \atop
{\scriptstyle \|\psi\| = 1}}} (\psi, H\psi)\ .
\end{equation}
Since $H$ depends on the coupling $v,$ so therefore does the
domain ${\cal D}(H).$ However, for the problems
considered,~either~$H$ has discrete eigenvalues, perhaps for $v$
greater than some critical coupling $v_1,$ or the entire spectrum
of $H$ is discrete~for $v>0.$ The kinetic potential $\bar{f}(s)$
associated with a given potential shape $f(r)$ is defined (for the
ground state $\psi$) by~a constrained minimization in which the
mean kinetic energy $s \equiv \langle-\Delta\rangle$ is kept
constant:
\begin{equation}\label{defKP}
\bar{f}(s) = \inf_{{{\scriptstyle \psi \in {\cal D}(H)} \atop
{\scriptstyle \|\psi\| = 1}} \atop {\scriptstyle (\psi,
-\Delta\psi) = s}} (\psi, f\psi)\ .
\end{equation}
The eigenvalue $F(v)$ of $H$ is then recovered from $\bar{f}(s)$
by a final minimization over $s$:
\begin{equation}\label{efroms}
F(v) = \min_{s>0}\!\left[s + v\,\bar{f}(s)\right].
\end{equation}
The spectral function $F(v)$ is concave ($F''(v) < 0$); moreover,
it has been shown \cite{inv1} that
\begin{equation}\label{convexities}
F''(v)\,\bar{f}''(s) = -\frac{1}{v^3}<0\ .
\end{equation}
Hence, $F(v)$ and $\bar{f}(s)$ have opposite convexities and are
related by the following Legendre transformations
$\bar{f}\leftrightarrow F$~\cite{GF}:
\begin{align}
&\bar{f}(s) = F'(v)\ ,\qquad s = F(v)-v\,F'(v)\label{legendre1}\
,\\&\frac{1}{v} = - \bar{f}'(s)\ ,\qquad \frac{F(v)}{v} =
\bar{f}(s) - s\,\bar{f}'(s)\label{legendre2}\ .
\end{align}
$F(v)$ is not necessarily monotone, but the kinetic potential
$\bar{f}(s)$ is monotone decreasing. Equation (\ref{legendre1})
enables us~also to use the coupling as a minimization parameter.
For this purpose, we write the coupling as $u$ and we have
from~\eq{efroms}
\begin{equation}\label{efromu}
F(v) = \min_{u>0}\!\left[F(u)-u\,F'(u) + v\,F'(u)\right].
\end{equation}
This is particularly useful in cases where $\bar{f}(s)$ is
difficult to find explicitly.

Another form of expression, useful for our present task, is
obtained if we change the kinetic-energy parameter from~$s$ to $r$
itself, by inverting the (monotone) function $\bar{f}(s)$ to
define the associated $K$-function by
\begin{equation}\label{Kfunction}
K^{[f]}(r) = s = \left(\bar{f}^{-1}\circ f\right)\!(r)\ .
\end{equation}
Now the energy formula Eq.~(\ref{efroms}) becomes
\begin{equation}\label{efromr}
F(v) = \min_{r>0}\!\left[K^{[f]}(r) + v\,f(r)\right].
\end{equation}
A sleight of hand may be perceived here since $K$ depends on $f.$
However, we do now have a relation that has $F$ on~one side and
$f$ on the other: our goal is to invert this expression, to effect
$F\rightarrow f.$ We shall do this below by constructing~a
sequence of approximate $K$-functions which do not depend on $f.$

%%%%%%%%%%%%%%%%%%%%%%%%%%%%%%%%%%%%%%%%%%%%%%%%%%%%%%%%%%%%%%%
\subsection{Smooth transformations and envelope approximations}
%%%%%%%%%%%%%%%%%%%%%%%%%%%%%%%%%%%%%%%%%%%%%%%%%%%%%%%%%%%%%%%
In this section, we consider potential shapes $f(r)$ that may be
written as smooth transformations $f(r) = g(h(r))$~of~a `basis
potential' $h(r)$. The idea is that we know the spectrum of
$-\Delta +v\,h(r)$ and we try to exploit this to study~the
spectrum of $-\Delta + v\,f(r).$ When the transformation function
$g$ has definite convexity ($g''$ does not change sign),~the
kinetic-potential formalism immediately allows us to derive energy
bounds. This is a consequence of Jensen's inequality
\cite{Jensen}, which may be expressed in our context by the
following:
\begin{eqnarray}\label{jineq}
\nonumber\mbox{$g$~is~convex}\qquad(g''\ge
0)\qquad&\Longrightarrow&\qquad(\psi,g(h)\psi)\ge g((\psi,h\psi))\
,\\\mbox{$g$~is~concave}\qquad(g''\le
0)\qquad&\Longrightarrow&\qquad(\psi,g(h)\psi)\le g((\psi,h\psi))\
.
\end{eqnarray}
More specifically, we have for the kinetic potentials
\begin{equation}\label{kpineq}
g''\ge0\qquad\Longrightarrow\qquad\bar{f}(s)\ge g(\bar{h}(s))\
;\qquad g''\le0\qquad\Longrightarrow\qquad\bar{f}(s)\le
g(\bar{h}(s))\ .
\end{equation}
We can summarize these results by writing $\bar{f}(s)\approx
g(\bar{h}(s))$ and remembering that the relation $\approx$
indicates an inequality whenever $g$ has definite convexity. The
expression of these results in terms of $K$-functions is even
simpler, for we have
\begin{equation}\label{kg}
K^{[f]} = \bar{f}^{-1}\circ f \approx
(g\circ\bar{h})^{-1}\circ(g\circ h)= \bar{h}^{-1}\circ h =
K^{[h]}\ .
\end{equation}
Thus, $K^{[f]}\approx K^{[h]}$ is the approximation we sought,
that no longer depends on $f.$ The corresponding energy bounds~are
provided by
\begin{equation}\label{eapprox}
E = F(v) \approx \min_{s > 0}\!\left[s +
v\,g\!\left(\bar{h}(s)\right)\right] =
\min_{r>0}\!\left[K^{(h)}(r) + v\,f(r)\right].
\end{equation}

%%%%%%%%%%%%%%%%%%%%%%%%%%%%%%%%%%%%%%%%%%%%%%%%%%%%%%%%%%%%%%%
\subsection{The envelope inversion sequence}\label{invseq}
%%%%%%%%%%%%%%%%%%%%%%%%%%%%%%%%%%%%%%%%%%%%%%%%%%%%%%%%%%%%%%%
We suppose that an eigenvalue $E$ of $H = -\Delta +v\,f(r)$ is known
as function $E = F(v)$ of the coupling parameter~$v > 0.$ In some
cases, such as the square well, the discrete eigenvalue may exist
only for sufficiently large coupling, $v > v_1.$~~The kinetic
potential $\bar{f}(s)$ may be obtained by inverting the Legendre
transformation in \eq{legendre1}. Thus
\begin{equation}\label{fbarfromF}
F(v) =
\min_{s>0}\!\left[s+v\,\bar{f}(s)\right]\qquad\leadsto\qquad
\bar{f}(s) =
\max_{v>v_1}\!\left[\frac{F(v)}{v}-\frac{s}{v}\right].
\end{equation}
We shall also need to invert the relation (\ref{efromr}) between
$F^{[n]}$ and $K^{[n]}$ by means of
\begin{equation}\label{KfromF}
K(r) = \max_{v>v_1}\!\left[F(v) - v\,f(r)\right].
\end{equation}

We begin with a seed potential shape $f^{[0]}(r)$ from which we
generate a sequence $\{f^{[n]}(r)\}_{n=0}^{\infty}$ of improving
potential approximations. The idea behind this sequence is that we
search for a transformation $g$ so that $g(f^{[n]}(r))$ is
close~to~$f(r)$ in the sense that the eigenvalue generated is
close to $F(v).$ The envelope approximation is used at each stage.
The~best transformation $g^{[n]}$ at stage $n$ is given by using
the current potential approximation $f^{[n]}(r)$ as an envelope
basis. We have:
\[
\bar{f} = g^{[n]}\circ\bar{f}^{[n]}\qquad\Longrightarrow\qquad
g^{[n]}=\bar{f}\circ\bar{f}^{[n]\,-1}\ .
\]
Thus$$f^{[n+1]}=g^{[n]}\circ f^{[n]}=\bar{f}\circ K^{[n]}\ .$$The
resulting inversion algorithm may be summarized by the following:
\medskip

\noindent{\bf inversion algorithm}
\begin{align}
f^{[n]}(r)\quad\longrightarrow\quad F^{[n]}(v)
\quad\longrightarrow\quad K^{[n]}(r)&=\max_{u >
v_1}\!\left[F^{[n]}(u) - u\,f^{[n]}(r)\right]\label{algor1},\\
f^{[n+1]}(r)&=\max_{v >
v_1}\!\left[\frac{F(v)}{v}-\frac{K^{[n]}(r)}{v}\right].\label{algor2}
\end{align}
The step $f^{[n]}(r)\longrightarrow F^{[n]}(v)$ is effected by
solving $\left(-\Delta + v\,f^{[n]}\right)\psi = E\,\psi$
numerically for $E = F^{[n]}(v).$

\subsection{Uniqueness}\label{unique}We consider now a singular
potential $f(r)$ of the form
\begin{equation}\label{formf}
f(r) = \frac{g(r)}{r}\ ,\qquad{\rm where}\qquad g(0) < 0\ ,\qquad
g'(r) \ge 0\ ,
\end{equation}
and $g(r)$ is not constant. Examples of this class of singular
potential shapes $f(r)$ are Yukawa $g(r) = -e^{-ar},$ Hulth\'en
$g(r) = -r/(e^{ar} -1),$ and linear-plus-Coulomb $g(r) = -a +
br^2,$ with $a,b > 0.$ With these assumptions, we have~proved~in
Ref.~\cite{FIS} the following

\nl\noindent{\bf Theorem~1} ~~~{\it The potential shape $f(r)$ in
$H = -\Delta + v\,f(r)$ is uniquely determined by the ground-state
energy~function $E = F(v).$}

%%%%%%%%%%%%%%%%%%%%%%%%%%%%%%%%%%%%%%%%%%%%%%%%%%%%%%%%%%%%%%%%%%%%%%%%%%%%%%%%%%%%%%%%
\section{Spectral data from Minkowski-space Bethe--Salpeter equation}\label{Sec:CKdata}
%%%%%%%%%%%%%%%%%%%%%%%%%%%%%%%%%%%%%%%%%%%%%%%%%%%%%%%%%%%%%%%%%%%%%%%%%%%%%%%%%%%%%%%%
\subsection{The raw data}
%%%%%%%%%%%%%%%%%%%%%%%%%%%%%
In Table~\ref{Tab:CK-combined}, we exhibit the binding energy $E$
versus coupling $v$ results from numerical solutions of the
Bethe--Salpeter equation for a system of two scalar particles each
of mass $m,$ bound by single or multiple exchange of a scalar
particle~of mass $\mu,$ computed by Carbonell and Karmanov in
Refs.~\cite[Table~1]{CK05a}, \cite[Table~1]{CK05b}, \cite[Tables~1
and 2]{CK05c}, and \cite[Table~1]{CK10c}. For comparison, we add
corresponding results of the Schr\"odinger equation with an
interaction potential of Yukawa form $V(r)=v\exp(-\mu\,r)/r,$
which, as shown in Appendix~\ref{ANRR}, constitutes the
nonrelativistic~limit~of~the~ladder~BS~quation. Interestingly, the
nonrelativistic binding energies emerging from the Schr\"odinger
equation with Yukawa potential seem to reproduce better the
ladder-plus-cross-ladder approximation findings than the ones from
mere ladder approximation.

\begin{table}[ht]\caption{Couplings $v$ and binding energies
$E$ arising from Bethe--Salpeter equations in either ladder or
ladder-plus-cross-ladder approximation for common mass $m=1$ of
the bound scalar bosons and mass $\mu=0.15$ or $\mu=0.5$ of the
exchanged~scalar~boson, or from the ladder-approximation
nonrelativistic limit, the Schr\"odinger equation with Yukawa
potential $V(r)=-v\exp(-\mu\,r)/r.$}\label{Tab:CK-combined}
\begin{center}\begin{tabular}{llclclr}\hline\hline\\[-1.5ex]
\multicolumn{6}{c}{$v$}&\multicolumn{1}{c}{$E$}
\\[1.5ex]\cline{1-6}\\[-1.5ex]\multicolumn{4}{c}{Minkowski-Space
Bethe--Salpeter Equation}&&\multicolumn{1}{c}{Schr\"odinger
Equation}\\[1.5ex]\cline{1-4}\cline{6-6}\\[-1.5ex]
\multicolumn{2}{c}{Ladder}&&\multicolumn{1}{c}{Ladder +
Cross-Ladder}&&\multicolumn{1}{c}{Yukawa Potential}
\\[1.5ex]\cline{1-2}\cline{4-4}\cline{6-6}\\[-1.5ex]
\multicolumn{1}{l}{$\mu=0.15$ \cite{CK05a,CK05c,CK10c}}&
\multicolumn{1}{l}{$\mu=0.50$ \cite{CK05a,CK05b,CK05c,CK10c}}&&
\multicolumn{1}{l}{$ \mu=0.50$ \cite{CK05b,CK05c}}&&
\multicolumn{1}{l}{$\mu=0.50$}\\[1.5ex]\hline\\[-1.5ex]
$0.5716$&$1.440$&&$1.21$&&1.034&$-0.01$\\
---&$2.01$&&$1.62$&&1.285&$-0.05$\\
$1.437$&$2.498$&&$1.93$&&1.532&$-0.10$\\
$2.100$&$3.251$&&$2.42$&&1.848&$-0.20$\\
$3.611$&$4.901$&&$3.47$&&2.204&$-0.50$\\
$5.315$&$6.712$&&$4.56$&&2.918&$-1.00$
\\[1.5ex]\hline\hline\end{tabular}\end{center}\end{table}

%%%%%%%%%%%%%%%%%%%%%%%%%%%%%%%%%%%%%%%%%%%
\subsection{Exchange-mass dependence}
\begin{figure}[ht]
\centering
\includegraphics[scale=1.355]{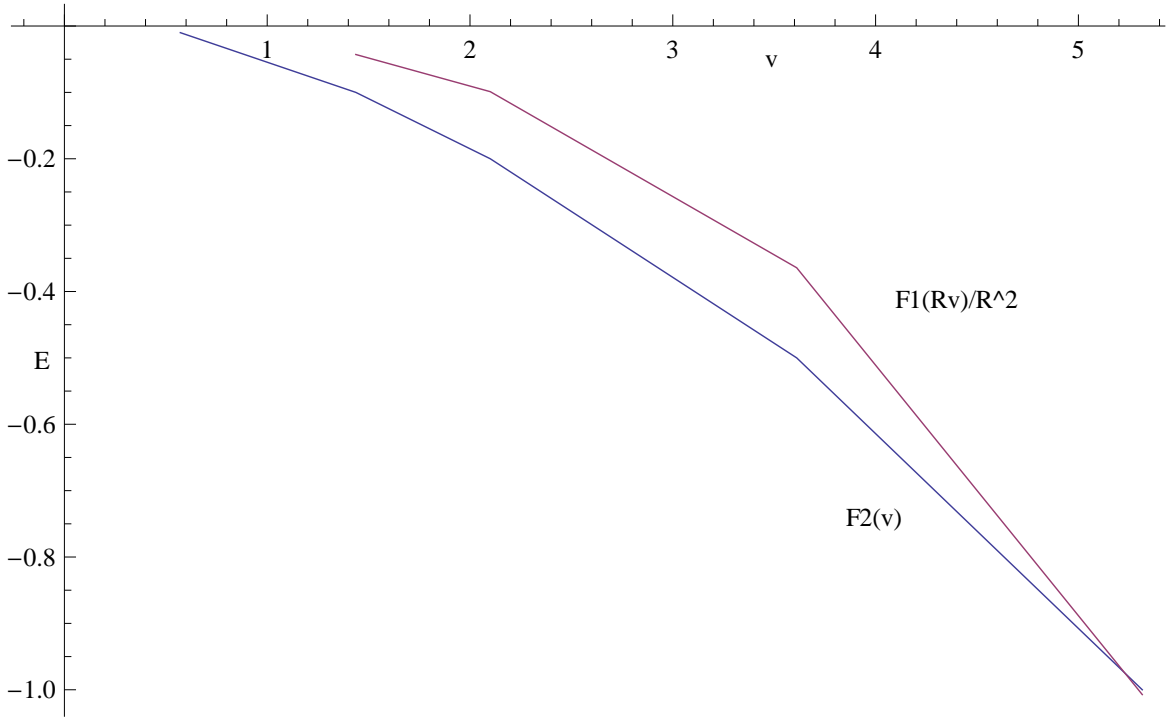}
\caption{Approximate exchange-mass dependence from scaling
arguments: Table \ref{Tab:CK-combined} data $F_2(v)$ vs.\ scaled
Table \ref{Tab:CK-combined} data $F_1(R\,v)/R^2.$}\label{figF2}
\end{figure}

%%%%%%%%%%%%%%%%%%%%%%%%%%%%%%%%%%%%%%%%%%%
In this subsection, we
demonstrate that the spectral data of Table~\ref{Tab:CK-combined}
for $\mu=0.5$ can be obtained approximately from the corresponding
data shown in Table~\ref{Tab:CK-combined} for $\mu=0.15$ by a
scale change in the potential of a Schr\"odinger model. This is
interesting if one expects to find that a potential such as the
Yukawa $V(r) = -v\,e^{-\mu\,r}/r$ would account approximately for
the spectral dependence of the problem on the exchange mass $\mu.$
Let us consider a Schr\"odinger operator given by
\begin{equation}\label{sromu}
 -\Delta + v\,\frac{f(r)}{r} \qquad\longrightarrow\qquad E = F(v)\ ,
\end{equation}
where $E$ is a discrete eigenvalue. A simple scaling argument
applied to the operator
\begin{equation}\label{sromu1}
H = -\frac{1}{2m}\,\Delta + v\,\frac{f(\mu\,r)}{r}
\end{equation}
shows that a corresponding discrete eigenvalue of $H$ is given in
terms of $F(v)$ by the formula
\begin{equation}\label{sroscale}
E = \frac{\mu^2}{2m}\,F\!\left(\frac{2m\,v}{\mu}\right).
\end{equation}
Thus, if we compare two different $\mu$ values, $\mu_1$ and
$\mu_2$ with $R\equiv\mu_1/\mu_2,$ and we write for the $\mu_1$
case $E_1 = F_1(v),$~then, under the scaling rule
(\ref{sroscale}), for the $\mu_2$ case we would have $E_2 =
F_1(R\,v)/R^2$. For our present problem, we have~$\mu_1=0.5,$
$\mu_2=0.15,$ and therefore $R = \mu_1/\mu_2 = 10/3.$ Hence, given
the second column of Table~\ref{Tab:CK-combined} expressed as $E =
E_1 = F_1(v)$, we would expect to generate data consistent with
the first column of Table~\ref{Tab:CK-combined} by the formula
$E_2 = (9/100)\,F_1(10v/3).$ Graphs of $F_2(v)$ and its
approximation in terms of the scaled $F_1(v)$ are shown in
Fig.~\ref{figF2}. The scaling law seems to~yield a rough
approximation.

%%%%%%%%%%%%%%%%%%%%%%%%%%%%%%%%%%%%%%%%%%%%%%%%%%%%%%%%%%%%%%%%%%%
\section{The construction of effective potentials}\label{Sec:EP}
\begin{figure}[ht]
\centering
\includegraphics[scale=.5456]{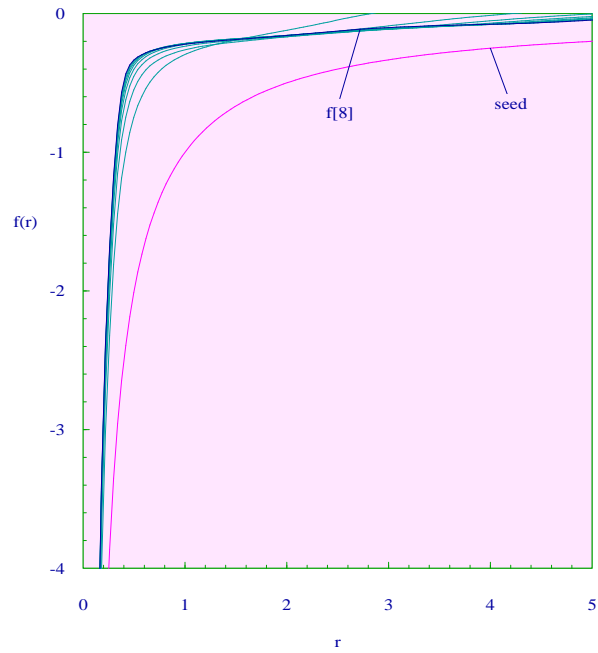}\\(a)\\
\includegraphics[scale=.5456]{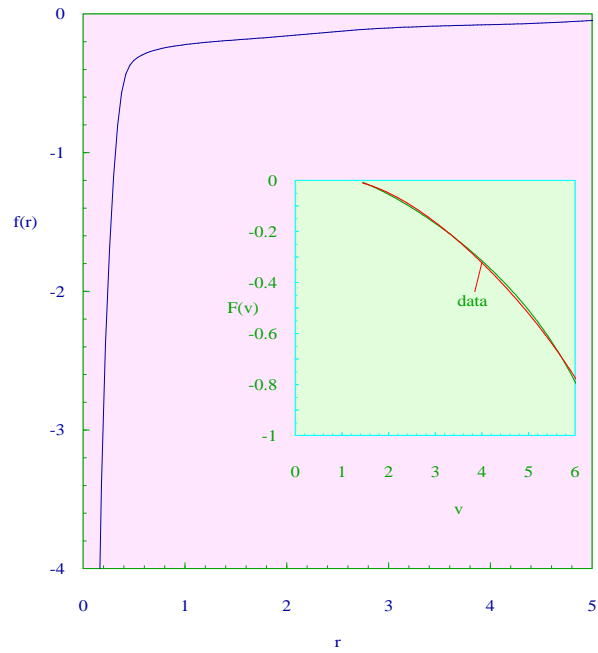}\\(b)
\caption{Geometric inversion of spectral data of bound states of
two scalar bosons described by a Minkowski-space Bethe--Salpeter
equation in ladder approximation for mass $\mu = 0.5$ of the
exchanged particle \cite{CK05a,CK05b,CK05c,CK10c}. Graph (a) shows
the first~eight~iterations $f[n],$ $n=1,2,\dots,8,$ starting from
the seed, while graph (b) shows the resulting potential shape
$f(r)$ and, in the inner graph,~the corresponding spectral curve
$F(v)$ along with the input data.}\label{figa}
\end{figure}

\begin{figure}[ht]
\centering
\includegraphics[scale=.5456]{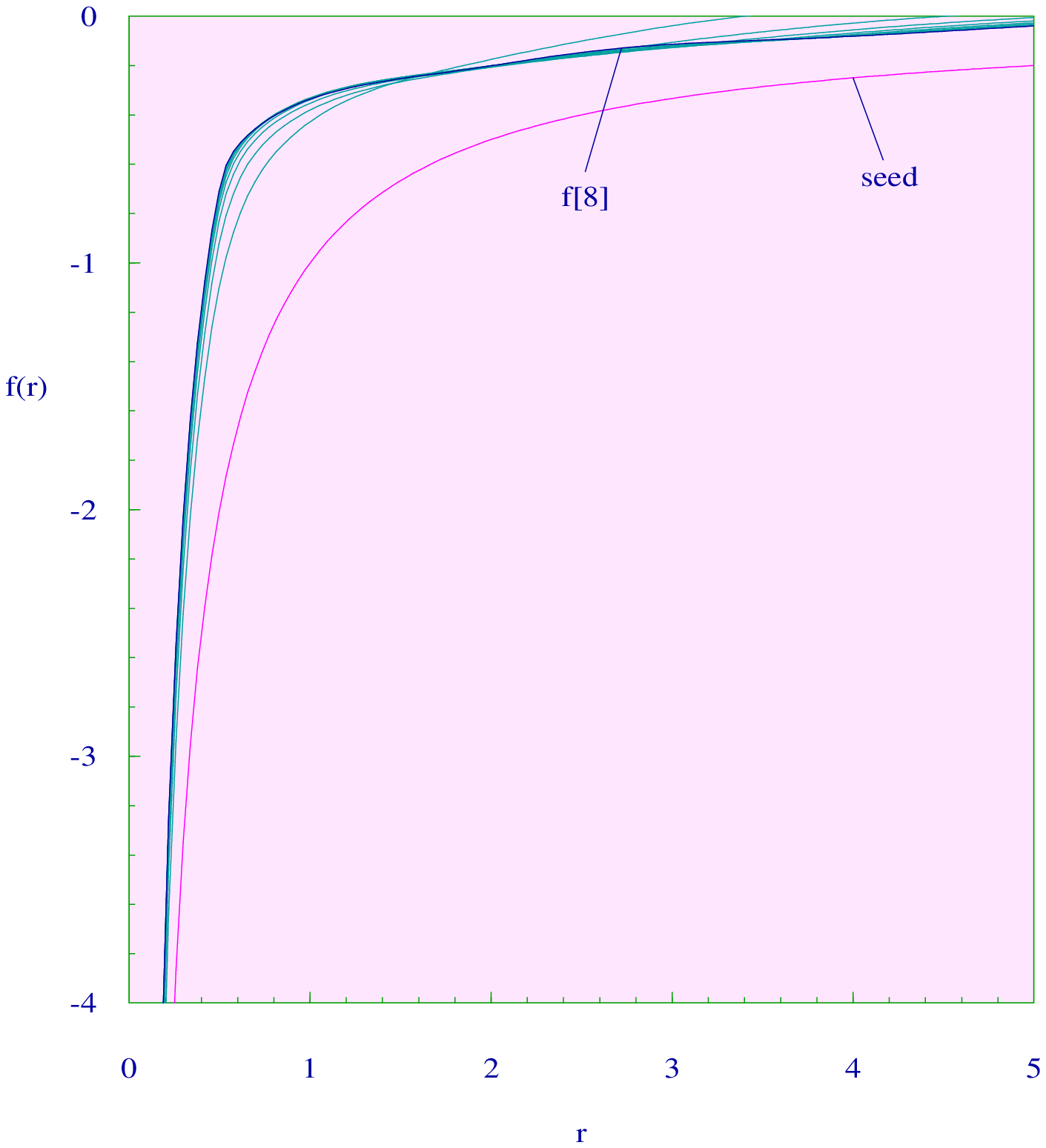}\\(a)\\
\includegraphics[scale=.5456]{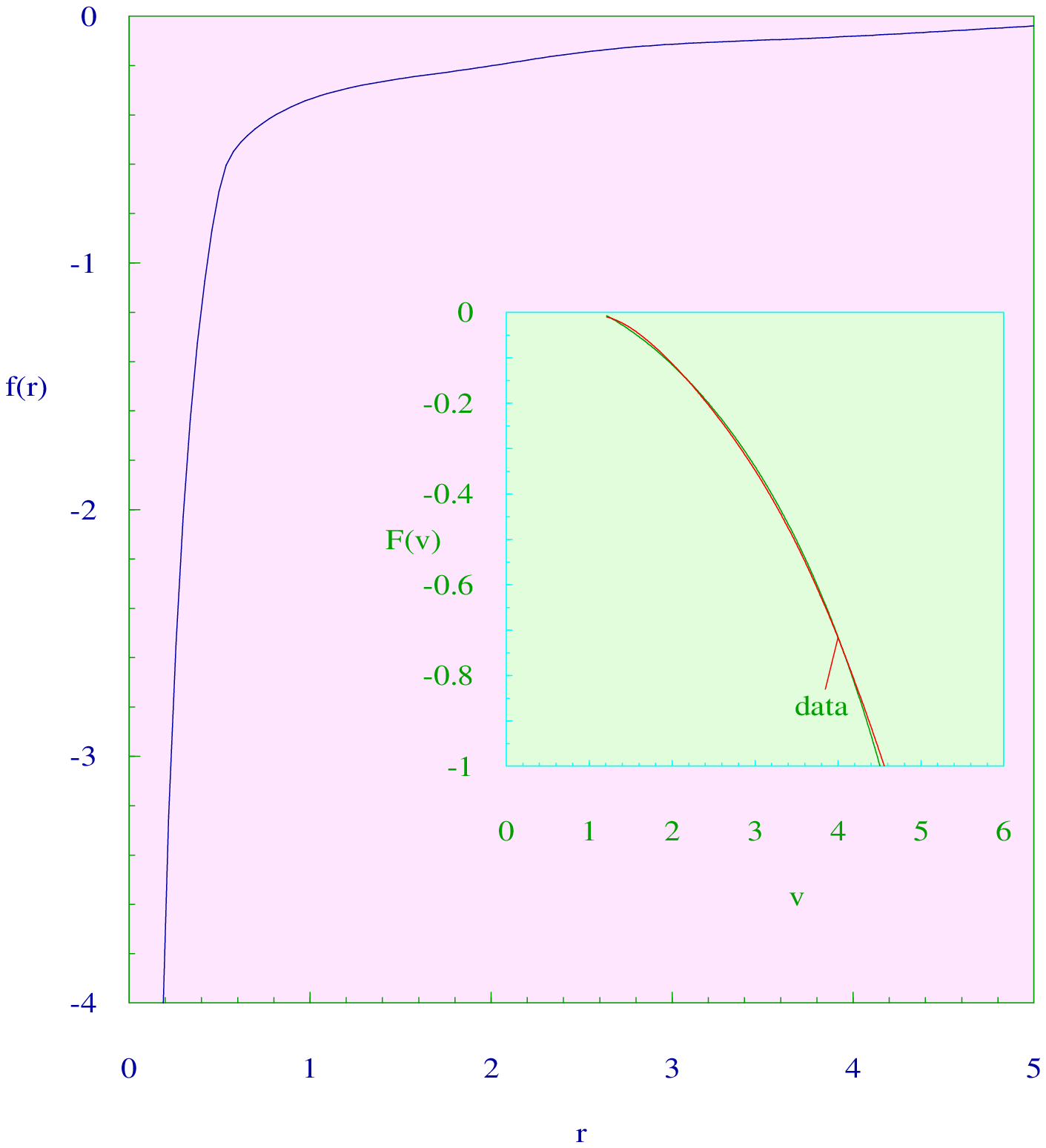}\\(b)
\caption{Geometric inversion of spectral data of bound states of
two scalar bosons described by a Minkowski-space Bethe--Salpeter
equation in ladder-plus-cross-ladder approximation for mass $\mu =
0.5$ of the exchanged particle \cite{CK05b,CK05c}. Graph
(a)~shows~the~first eight iterations $f[n],$ $n=1,2,\dots,8,$
starting from the seed, while graph (b) shows the resulting
potential shape $f(r)$ and, in~the inner graph, the corresponding
spectral curve $F(v)$ along with the input data.}\label{figb}
\end{figure}

\begin{figure}[ht]
\centering
\includegraphics[scale=.5456]{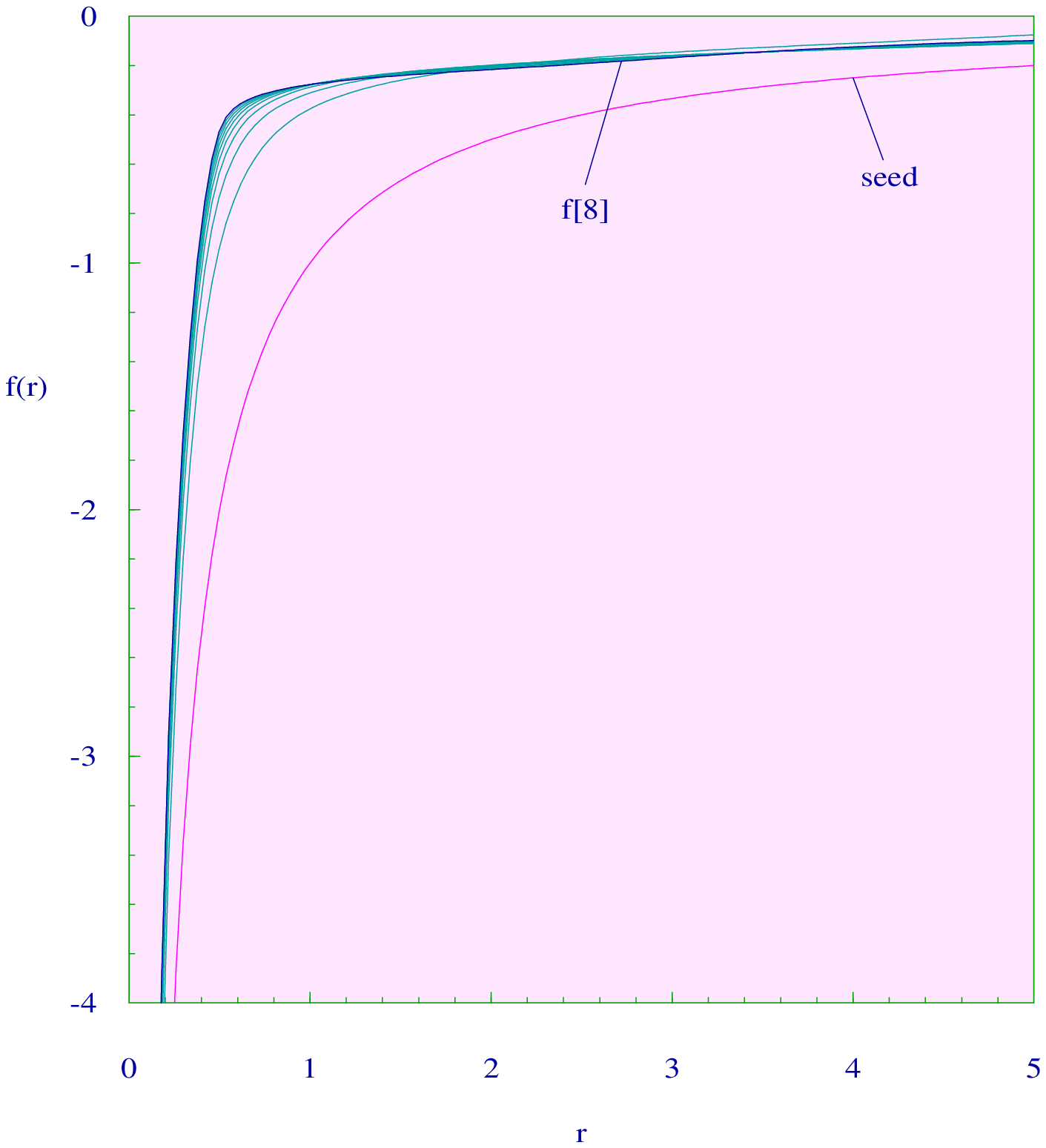}\\(a)\\
\includegraphics[scale=.5456]{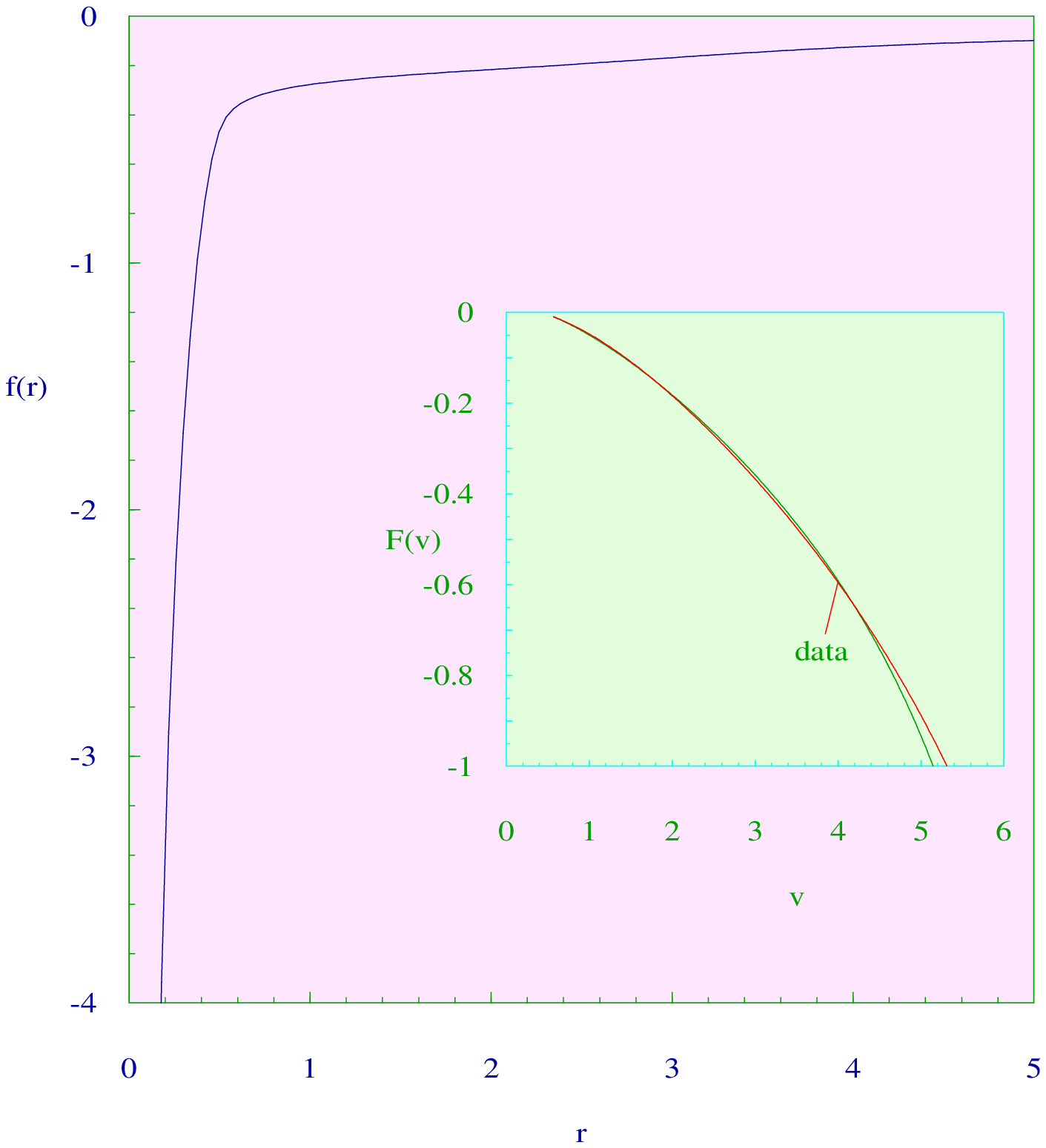}\\(b)
\caption{Geometric inversion of spectral data of bound states of
two scalar bosons described by a Minkowski-space Bethe--Salpeter
equation in ladder approximation for mass $\mu = 0.15$ of the
exchanged particle \cite{CK05a,CK05c,CK10c}. Graph (a) shows
the~first~eight~iterations $f[n],$ $n=1,2,\dots,8,$ starting from
the seed, while graph (b) shows the resulting potential shape
$f(r)$ and, in the inner graph,~the corresponding spectral curve
$F(v)$ along with the input data.}\label{figc1}
\end{figure}

\begin{figure}[ht]
\centering
\includegraphics[scale=.5456]{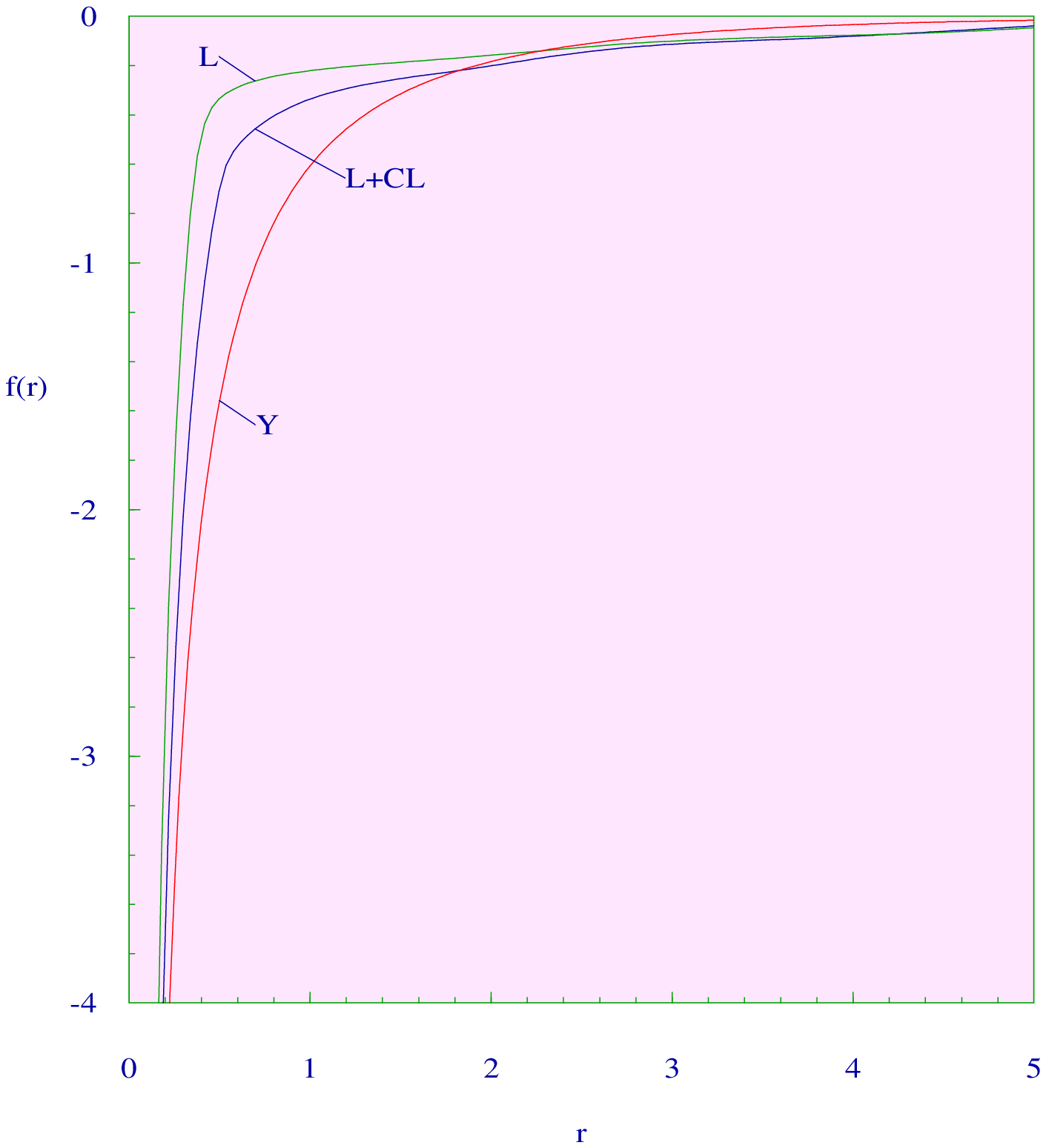}
\caption{Potential shapes $f(r)$ from Figs.~\ref{figa} and
\ref{figb} resulting from geometric inversion of the ladder (L)
and ladder-plus-cross-ladder (L + CL) Bethe--Salpeter findings
\cite{CK05a,CK05b,CK05c,CK10c} for the masses of two-scalar-boson
bound states, with exchanged-boson~mass~$\mu = 0.5$, compared with
the corresponding Yukawa potential $Y(r) = -\exp(-\mu\,r)/r,$
arising in the nonrelativistic limit of the
ladder~case.}\label{figy}
\end{figure}

%%%%%%%%%%%%%%%%%%%%%%%%%%%%%%%%%%%%%%%%%%%%%%%%%%%%%%%%%%%%%%%%%%%
We now consider the BS spectral data collected in
Table~\ref{Tab:CK-combined} and we use our inversion theory
\cite{FIS} to answer the~question what potential shape $f(r)$ in
the Schr\"odinger Hamiltonian $H = -1/(2m)\,\Delta + v\,f(r)$
would generate the corresponding binding energies $E$ for the
given values of the coupling parameter $v$? For this purpose, we
take $m = m_1\,m_2/(m_1+m_2)$ with $m_1=m_2 = 1.$ We adopt the
{\it inversion algorithm} of \eq{algor2} with the pure Coulomb
seed potential $f^{[0]}(r) = -1/r.$ For the three sets of BS data
in Table \ref{Tab:CK-combined}, our results are exhibited
respectively in Figs.~\ref{figa}, \ref{figb} and \ref{figc1}. In
each case, we first show a sequence of eight iterations and then
depict the last potential iteration, $f^{[8]}(r),$ along with the
eigenvalue curve $F(v)$ of the Bethe--Salpeter data and the
corresponding eigenvalue curve of the Hamiltonian $H = -\Delta +
v\,f^{[8]}(r).$ For comparison, Fig.~\ref{figy} shows the
reconstructed potentials for both ladder (L) and
ladder-plus-cross-ladder (L + CL) kernels with an
exchanged-particle mass $\mu =0.5,$ along with the corresponding
Yukawa-potential shape $Y(r)\equiv-\exp(-\mu\,r)/r$. Judged by the
eye, the Yukawa potential seems to be closer to the inversion
output for the ladder-plus-cross-ladder~case than for the mere
ladder case. In view of the ordering of couplings $v$ in Table
\ref{Tab:CK-combined}, this observation is no genuine surprise.

%%%%%%%%%%%%%%%%%%%%%%%%%%%%%%%%%%%%%%%%%%%
\section{Form factors}\label{Sec:FF}
\begin{figure}[ht]
\centering
\includegraphics[scale=.565]{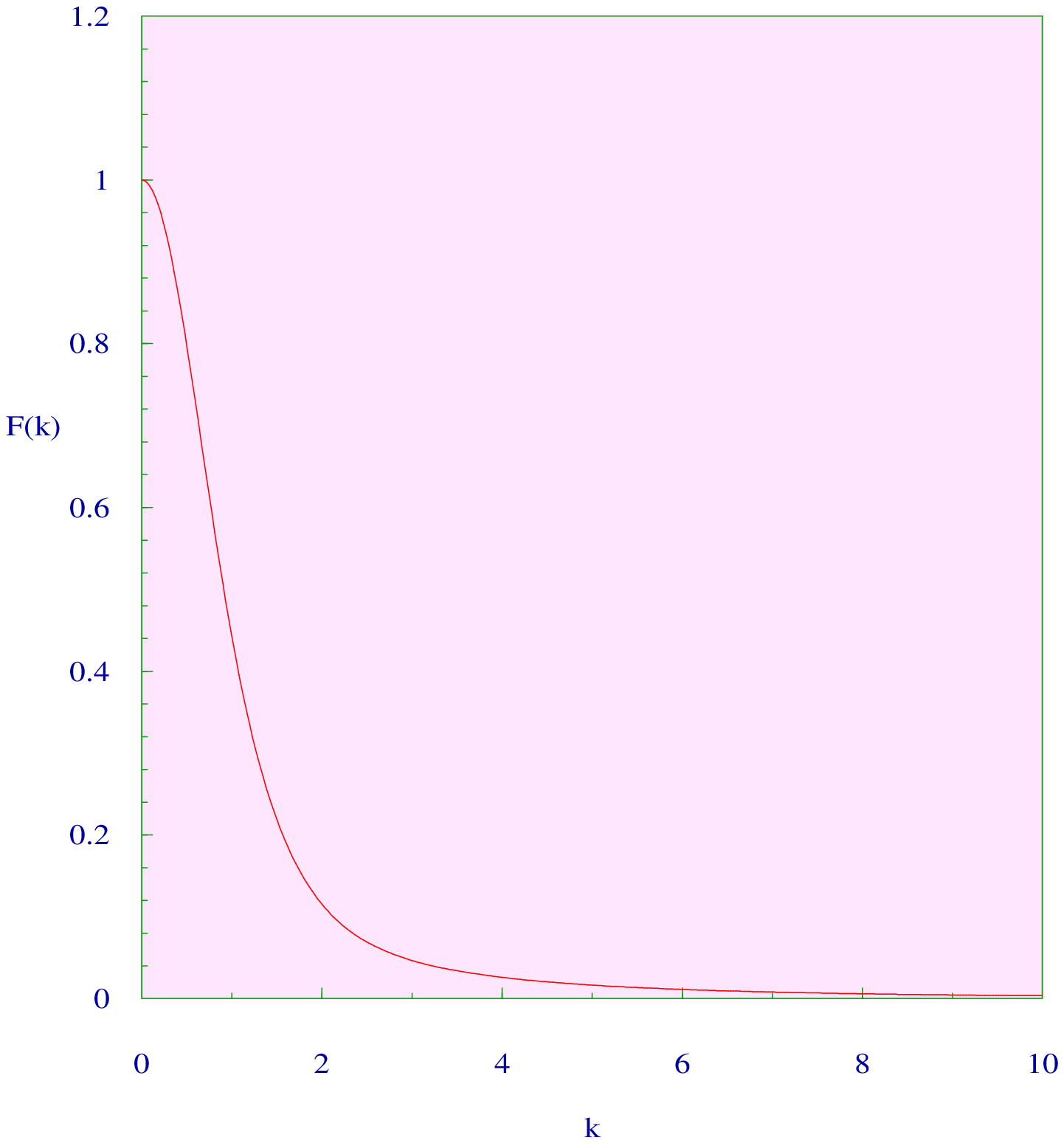}\\(a)\\
\includegraphics[scale=.565]{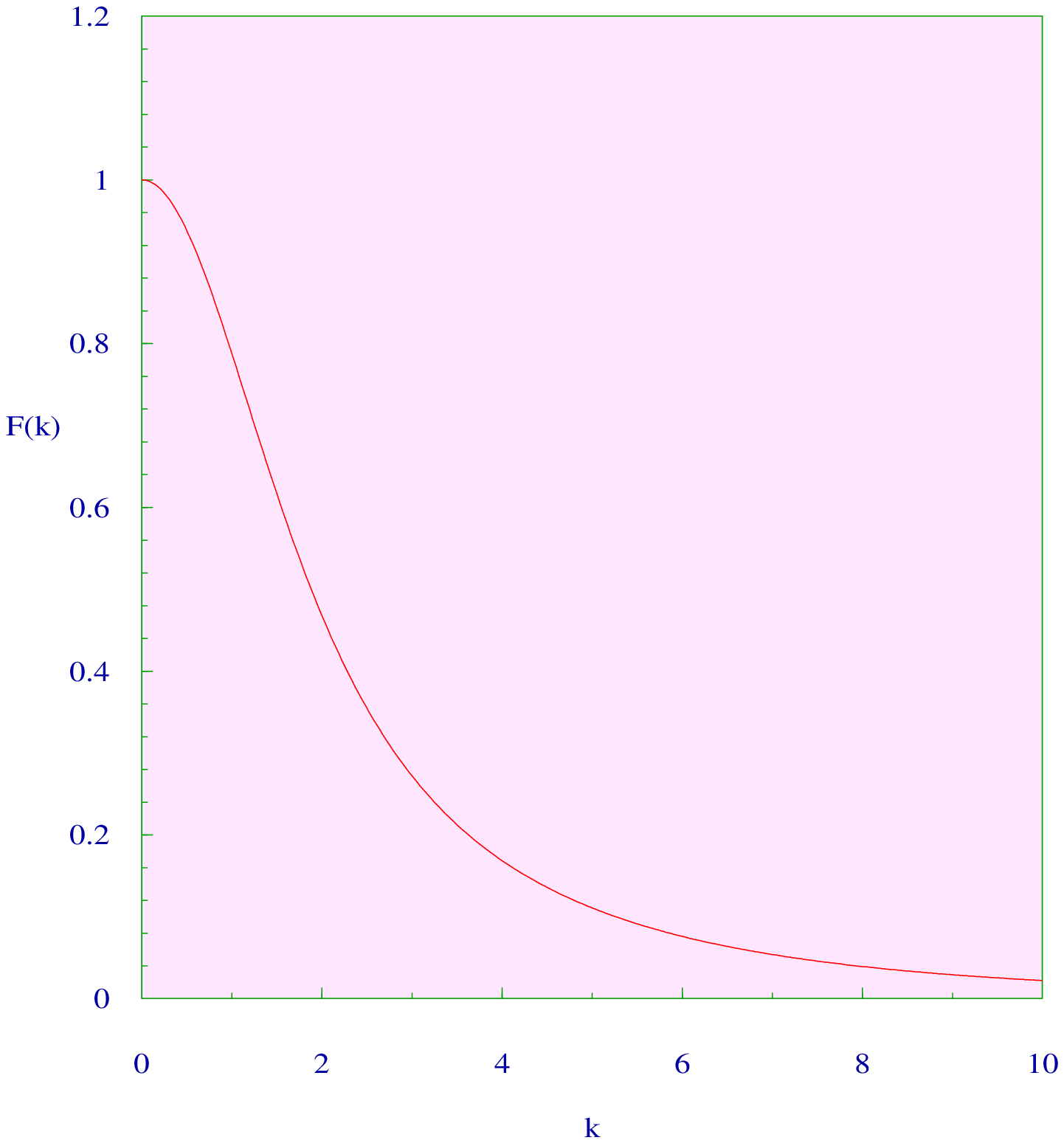}\\(b)
\caption{Momentum-space form factor $F(k)$ of the ground-state
Schr\"odinger solution mimicking the results of the
Bethe--Salpeter equation in ladder (a) and
ladder-plus-cross-ladder (b) approximation with coupling $v = 5$
and exchanged-particle mass~$\mu = 0.5.$}\label{figfLCL50}
\end{figure}

\begin{figure}[ht]
\centering
\includegraphics[scale=.565]{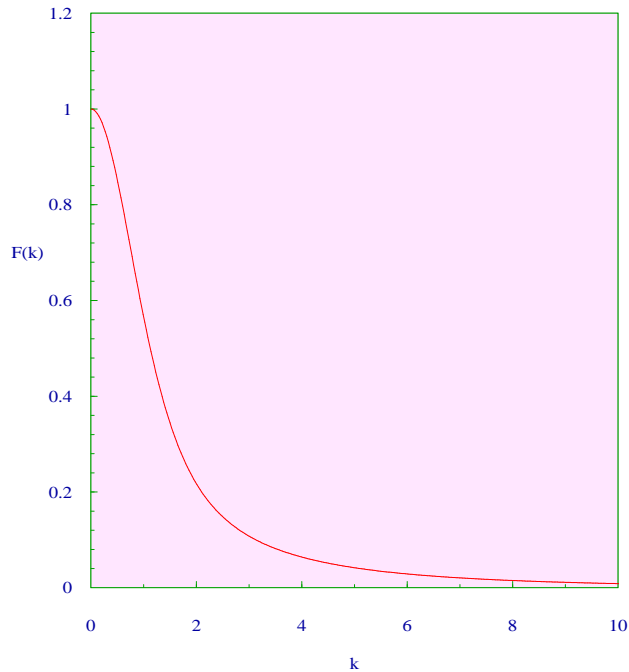}
\caption{Momentum-space form factor $F(k)$ of the ground-state
Schr\"odinger solution mimicking the results of the
Bethe--Salpeter equation in ladder approximation with coupling $v
= 5$ and exchanged-particle mass~$\mu = 0.15.$}\label{figfL05}
\end{figure}

%%%%%%%%%%%%%%%%%%%%%%%%%%%%%%%%%%%%%%%%%%%
In quantum mechanics, the three-dimensional form factor
$F(\bm{k})$ of a bound state described by its configuration-space
wave function $\psi(\bm{x})$ is nothing else but the Fourier
transform of the corresponding charge density
$\rho(\bm{x})\equiv|\psi(\bm{x})|^2.$ For a given coupling
strength $v$ and common mass $m$ of the two bound-state
constituents, the reduced radial wave functions $u(r)$ of
$s$-states satisfy an ordinary differential equation that
determines the associated binding-energy eigenvalues $E$:
\begin{equation}\label{radialde}
-\frac{1}{m}\,u''(r)+v\,f(r)\,u(r) = E\,u(r)\ ,\qquad u(0) = 0\ .
\end{equation}
It proves convenient to normalize the radial wave functions $u(r)$
such that $\int_0^{\infty}{\rm d}r\,|u(r)|^2 = 1.$ In momentum
space,~any corresponding form factor $F(k)$ is given by the
Fourier--Bessel transform of the radial density $u^2(r)$, that is
to say,~by
\begin{equation}\label{ffactor}
F(k) = \frac{1}{k}\int\limits_0^{\infty}{\rm
d}r\,\frac{\sin(k\,r)}{r}\,u^2(r)\ ,\qquad F(0)=1\ .
\end{equation}
Figures~\ref{figfLCL50} and \ref{figfL05} depict the ground-state
form factors $F(k)$ for $v = 5$ and $m = 1$ for each of the three
potential shapes $f(r)$ shown in Figs.~\ref{figa}--\ref{figc1},
obtained by application of our geometrical inversion technique to
the spectral data of Table~\ref{Tab:CK-combined}. By comparing
Figs.~\ref{figfLCL50}(a) and \ref{figfL05}, we see that the
quantum-mechanical form factors broaden with increasing mass $\mu$
of the exchange particle. Likewise, a comparison of the two plots
in Fig.~\ref{figfLCL50} reveals that the quantum-mechanical form
factors broaden when taking into account higher-order corrections
in the BS interaction kernel, that is to say, when working in the
somewhat more sophisticated ladder-plus-cross-ladder approximation
instead of in the na\"ive~ladder approximation. These form factors
$F(k)$ constitute the three-dimensional counterparts of the
four-dimensional form factor $F(Q^2)$ found from the
ladder-plus-cross-ladder Minkowski-space BS equation with
exchanged-boson mass $\mu=0.5$ in
Refs.~\cite{CK06,CK07,CK08,CK10c}.

\clearpage

%%%%%%%%%%%%%%%%%%%%%%%%%%%
\section{Conclusion}
%%%%%%%%%%%%%%%%%%%%%%%%%%%
For a {\em quantum-mechanical\/} description of relativistic
systems in which a single particle is bound to a fixed center,~one
uses the Klein--Gordon equation or the Dirac equation. For the
analysis of systems composed of more than one particle within {\em
quantum field theory\/}, the Bethe--Salpeter formalism is
required. In a series of papers, Carbonell {\em et al.} discussed
the bound states of a system of two charged scalar bosons by means
of the Bethe--Salpeter equation in Minkowski-space representation,
in contrast to most studies of this kind which rely on the
Euclidean-space formulation of this~equation. These investigations
report some numerical results for the binding energies, in a
variety of cases, as functions $E(v)$~of~a coupling parameter $v.$
What we have done in the present analysis is to employ a
geometric~spectral~inversion theory~to reconstruct, in each case,
the potential shape $f(r)$ in a Schr\"odinger model $H =
-\Delta/(2m) + v\,f(r)$ which would~have~the same energy curve
$E(v).$ As more complete quantum-field-theoretic spectral data
becomes available, we shall be able~to reveal more details of such
spectrally-equivalent potential models.

% ------------------------------------------------------
\section*{Acknowledgments}
% ------------------------------------------------------
One of us (RLH) gratefully acknowledges both partial financial
support of this research under Grant No.\ GP3438 from the Natural
Sciences and Engineering Research Council of Canada and the
hospitality of the Institute for High Energy Physics of the
Austrian Academy of Sciences, Vienna, where part of the work was
done.\medskip

\newpage

\appendix
\section{Nonrelativistic Reduction of the Bethe--Salpeter Equation}
\label{ANRR}For the sake of completeness we briefly sketch how, by
successive application of a sequence of simplifying assumptions
and approximations, the Bethe--Salpeter equation for two
bound-state constituents both of spin zero may be reduced to an
equation of motion of Schr\"odinger form with all interactions
represented by a static potential. Regarding~{\em kinematics}, for
a bound state of two particles discriminated by a label $i=1,2$
the relation between total momentum $P$ and relative momentum $p$
of the constituents, on the one hand, and the individual-particle
momenta $p_1,$ $p_2,$ on the other hand,~reads$$P\equiv p_1+p_2\
,\qquad
p\equiv\eta_2\,p_1-\eta_1\,p_2\qquad\Longleftrightarrow\qquad
p_1=\eta_1\,P+p\ ,\qquad p_2=\eta_2\,P-p\ ,$$where $\eta_1,$
$\eta_2$ denote two real parameters satisfying $\eta_1+\eta_2=1.$
In the center-of-momentum frame of some bound~state~of mass
$M=\sqrt{P^2},$ defined by $\bm{P}\equiv\bm{p}_1+\bm{p}_2=\bm{0}$
and therefore $P=(M,\bm{0}),$ the individual-particle momenta
$p_1,$ $p_2,$~become $$p_1^0=\eta_1\,M+p^0\ ,\qquad
p_2^0=\eta_2\,M-p^0\ ,\qquad \bm{p}_1=\bm{p}\
,\qquad\bm{p}_2=-\bm{p}\ .$$

Our starting point of the nonrelativistic reduction is the
Bethe--Salpeter equation in momentum-space representation
\begin{equation}\Phi(p,P)=\frac{{\rm i}}{(2\pi)^4}\,S_1(p_1)\int{\rm
d}^4q\,K(p,q,P)\,\Phi(q,P)\,S_2(-p_2)\ .\label{ABSE}\end{equation}
Its {\em instantaneous approximation\/} assumes that in the
center-of-momentum frame of the bound state the Bethe--Salpeter
interaction kernel $K(p,q,P)$ depends only on the (initial and
final) spatial relative momenta $\bm{p},\bm{q}$:
$K(p,q,P)=K(\bm{p},\bm{q}).$ Then, integrating over $p_0$ reduces
the Bethe--Salpeter equation (\ref{ABSE}) to a kind of
instantaneous Bethe--Salpeter equation$$\phi(\bm{p})=\frac{{\rm
i}}{2\pi}\int{\rm d}p_0\,S_1(p_1)\,S_2(-p_2)\int\frac{{\rm
d}^3q}{(2\pi)^3}\,K(\bm{p},\bm{q})\,\phi(\bm{q})$$for the
`Salpeter amplitude' $\phi(\bm{p}),$ defined as integral of
$\Phi(p,P)$ over the time component $p_0$ of the relative
momentum~$p$:$$\phi(\bm{p})\equiv\frac{1}{2\pi}\int{\rm
d}p_0\,\Phi(p,P)\ .$$Replacing the propagator $S_i(p)$ of
bound-state constituent $i$ by its free counterpart $S_i^{(0)}(p)$
entails the Salpeter equation
\begin{equation}\phi(\bm{p})=\frac{{\rm i}}{2\pi}\int{\rm
d}p_0\,S_1^{(0)}(p_1)\,S_2^{(0)}(-p_2)\int\frac{{\rm
d}^3q}{(2\pi)^3}\,K(\bm{p},\bm{q})\,\phi(\bm{q})\
.\label{ASE}\end{equation}

The free Feynman propagator $S_i^{(0)}(p_i)$ in momentum space of
a {\em scalar boson\/} of mass $m_i$ and momentum $p_i$
is~given~by $$S_i^{(0)}(p_i)=S_i^{(0)}(-p_i)=\frac{{\rm
i}}{p_i^2-m_i^2+{\rm i}\,\varepsilon}\
,\qquad\varepsilon\downarrow0\ ,\qquad i=1,2\ .$$In terms of
relativistic free-particle energies
$E_i(\bm{p})\equiv\sqrt{\bm{p}^2+m_i^2}$ the unique\footnote{A
different decomposition of the free scalar-boson propagator
$S_i^{(0)}(p_i)$ that is, however, not compatible with the Cauchy
residue theorem~is$$S_i^{(0)}(p_i)=\frac{{\rm
i}}{2\,p_i^0}\left[\frac{1}{p_i^0-E_i(\bm{p})+{\rm
i}\,\varepsilon}+\frac{1}{p_i^0+E_i(\bm{p})-{\rm
i}\,\varepsilon}\right].$$} {\em partial fraction decomposition\/}
of $S_i^{(0)}(p_i)$~is$$S_i^{(0)}(p_i)=\frac{{\rm
i}}{2\,E_i(\bm{p})}\left[\frac{1}{p_i^0-E_i(\bm{p})+{\rm
i}\,\varepsilon}-\frac{1}{p_i^0+E_i(\bm{p})-{\rm
i}\,\varepsilon}\right],\qquad\varepsilon\downarrow0\ ,\qquad
i=1,2\ .$$Evaluating the integral over the two propagators in
Eq.~(\ref{ASE}) by contour integration and Cauchy's residue
theorem~gives$$\int{\rm d}p_0\,S_1^{(0)}(p_1)\,S_2^{(0)}(-p_2)
=\frac{2\pi\,{\rm i}}{4\,E_1(\bm{p})\,E_2(\bm{p})}
\left[\frac{1}{M-E_1(\bm{p})-E_2(\bm{p})}
-\frac{1}{M+E_1(\bm{p})+E_2(\bm{p})}\right].$$

The three-dimensional reduction of the relativistically covariant
Bethe--Salpeter equation (\ref{ABSE}) to a Schr\"odinger-type
equation is applicable to nearly nonrelativistic and {\em weakly
bound\/} states composed of sufficiently heavy constituents. For
such systems, the first contribution to the above integral over
propagators may be assumed to dominate the~second~one,\newpage
$$\frac{1}{M-E_1(\bm{p})-E_2(\bm{p})}
\gg\frac{1}{M+E_1(\bm{p})+E_2(\bm{p})}\ ,$$since in this situation
$M\approx E_1(\bm{p})+E_2(\bm{p}),$ such that the second term in
the propagator integral may be safely~neglected:$$\int{\rm
d}p_0\,S_1^{(0)}(p_1)\,S_2^{(0)}(-p_2) \approx\frac{2\pi\,{\rm
i}}{4\,E_1(\bm{p})\,E_2(\bm{p})}\,\frac{1}{M-E_1(\bm{p})-E_2(\bm{p})}\
.$$Adopting this standard approximation, one arrives at the
reduced Salpeter equation for spin-0 bound-state constituents
\begin{equation}\left[E_1(\bm{p})+E_2(\bm{p})\right]\phi(\bm{p})
-\frac{1}{4\,E_1(\bm{p})\,E_2(\bm{p})}\int\frac{{\rm
d}^3q}{(2\pi)^3}\,K(\bm{p},\bm{q})\,\phi(\bm{q})=M\,\phi(\bm{p})\
.\label{ARSE}\end{equation}Moreover, assuming that, in the first
term of the propagator integral, the factor
$[M-E_1(\bm{p})-E_2(\bm{p})]^{-1}$ varies faster than the factor
$[E_1(\bm{p})\,E_2(\bm{p})]^{-1}$ justifies the substitution
$E_i(\bm{p})\approx m_i$ in the denominator of the {\em
interaction~term\/}~in~Eq.~(\ref{ARSE}). Finally, the
nonrelativistic expansion $E_i(\bm{p})\approx
m_i+\bm{p}^2/(2\,m_i)$ of the free energies leads to the
Schr\"odinger-type equation\begin{equation}
\left(m_1+m_2+\frac{\bm{p}^2}{2\,m_1}+\frac{\bm{p}^2}{2\,m_2}\right)
\phi(\bm{p})-\frac{1}{4\,m_1\,m_2}\int\frac{{\rm
d}^3q}{(2\pi)^3}\,K(\bm{p},\bm{q})\,\phi(\bm{q})=M\,\phi(\bm{p})\
.\label{ANRL}\end{equation}

The Schr\"odinger equation governing the dynamics of two
particles, of masses $m_1,m_2,$ interacting via a potential
$V(\bm{x})$ involving their relative coordinate
$\bm{x}\equiv\bm{x}_1-\bm{x}_2,$ in a bound state of mass $M$
reads, in configuration-space~representation,
$$\left[m_1+m_2-\frac{\Delta_x}{2\,m_1}-\frac{\Delta_x}{2\,m_2}
+V(\bm{x}) \right]\psi(\bm{x})=M\,\psi(\bm{x})\ .$$Upon
introduction, for configuration-space wave function $\psi(\bm{x})$
and interaction potential $V(\bm{x}),$ their Fourier transforms
$$\widetilde\psi(\bm{p})=\int\frac{{\rm d}^3x}{(2\pi)^{3/2}}\,{\rm
e}^{-{\rm i}\,\bm{p}\cdot\bm{x}}\,\psi(\bm{x})\ ,\qquad\widetilde
V(\bm{p})=\int{\rm d}^3x\,{\rm e}^{-{\rm
i}\,\bm{p}\cdot\bm{x}}\,V(\bm{x})\ ,$$the latter proving to be a
very convenient choice, the Schr\"odinger equation becomes in
momentum-space representation\begin{equation}
\left(m_1+m_2+\frac{\bm{p}^2}{2\,m_1}+\frac{\bm{p}^2}{2\,m_2}\right)
\widetilde\psi(\bm{p})+\int\frac{{\rm d}^3q}{(2\pi)^3}\,\widetilde
V(\bm{p}-\bm{q})\,\widetilde\psi(\bm{q})=M\,\widetilde\psi(\bm{p})
\label{ASchEm}\ .\end{equation}Assuming the kernel
$K(\bm{p},\bm{q})$ to be of convolution type, i.e.,
$K(\bm{p},\bm{q})=K(\bm{p}-\bm{q}),$ the comparison of the
reduced~Salpeter equation in nonrelativistic limit (\ref{ANRL})
with the momentum-space Schr\"odinger equation (\ref{ASchEm})
allows for the identification$$\widetilde
V(\bm{p}-\bm{q})=-\frac{1}{4\,m_1\,m_2}\,K(\bm{p}-\bm{q})
\qquad\Longleftrightarrow\qquad\widetilde
V(\bm{p})=-\frac{1}{4\,m_1\,m_2}\,K(\bm{p})\ .$$

As an illustration of this relationship, let us demonstrate how
the {\em Yukawa potential\/} arises from {\em single-boson
exchange\/} between our two bound-state constituents $i=1,2.$ Let
$g_i$ denote the interaction strength (having the mass dimension
1) of the three-boson coupling of the spin-0 bound-state
constituent $i$ (of mass $m_i$) to some spin-0 force mediator~of
mass $\mu.$ Introducing the momentum transfer $k\equiv p-q,$ the
resulting one-boson exchange contribution to the
interaction~kernel~is$${\rm i}\,K(p,q,P)={\rm i}\,K(k)=\frac{{\rm
i}\,({\rm i}\,g_1)\,({\rm i}\,g_2)}{k^2-\mu^2}\ ;$$contenting
oneself with this form entails the {\em ladder approximation\/} to
Eq.~(\ref{ABSE}). In instantaneous limit, this kernel~reads
$$K(\bm{k})=\frac{g_1\,g_2}{\bm{k}^2+\mu^2}\ .$$As consequence of
the spherical symmetry of $K(\bm{k}),$ the Fourier transformation
of $\widetilde V(\bm{k})$ yields the spherically symmetric
configuration-space {\em Yukawa potential\/} $V(\bm{x})=V(r),$
$r\equiv|\bm{x}|.$ For convenience, we represent it in the form
$V(r)\equiv v\,f(r)$:$$V(\bm{x})=\int\frac{{\rm
d}^3k}{(2\pi)^3}\,{\rm e}^{{\rm i}\,\bm{k}\cdot\bm{x}}\,\widetilde
V(\bm{k})=-\int\frac{{\rm d}^3k}{(2\pi)^3}\,\frac{{\rm e}^{{\rm
i}\,\bm{k}\cdot\bm{x}}\,K(\bm{k})}{4\,m_1\,m_2}
=-\frac{g_1\,g_2}{16\pi\,m_1\,m_2}\,\frac{{\rm e}^{-\mu\,r}}{r}
=V(r)\equiv v\,f(r)\ ,\qquad f(r)=-\frac{{\rm e}^{-\mu\,r}}{r}\
.$$Hence our coupling constant $v$ is related to the mass and
coupling parameters of the underlying quantum field theory~by
$$v=\frac{g_1\,g_2}{16\pi\,m_1\,m_2}\ .$$Especially, for {\em
identical\/} bound-state constituents, clearly satisfying
$m_1=m_2=m$ and $g_1=g_2=g,$ this result~becomes
$$v=\frac{g^2}{16\pi\,m^2}\ .$$\newpage

%%%%%%%%%%%%%%%%%%%%%%%%%%%%%%%%%%%


\begin{thebibliography}{99}
%%%%%%%%%%%%%%%%%%%%%%%%%%%%%%%%%%%

% Bethe-Salpeter formalism
\bibitem{BSE_APS}H.~A.~Bethe and E.~E.~Salpeter, Phys.~Rev.~{\bf 82}
(1951) 309.
\bibitem{GML}M.~Gell-Mann and F.~Low, Phys.~Rev.~{\bf 84} (1951)
350.
\bibitem{BSE}E.~E.~Salpeter and H.~A.~Bethe, Phys.~Rev.~{\bf 84}
(1951) 1232.

% Wick
\bibitem{Wick}G.~C.~Wick, Phys.~Rev.~{\bf 96} (1954) 1124.

% Related CKMB papers
\bibitem{CK05a}V.~A.~Karmanov and J.~Carbonell, Eur.~Phys.~J.~A
{\bf 27} (2006) 1, arXiv:hep-th/0505261.
\bibitem{CK05b}J.~Carbonell and V.~A.~Karmanov, Eur.~Phys.~J.~A
{\bf 27} (2006) 11, arXiv:hep-th/0505262.
\bibitem{CK05c}V.~A.~Karmanov and J.~Carbonell,
Nucl.~Phys.~Proc.~Suppl.~{\bf 161} (2006) 123,
arXiv:nucl-th/0510051.
\bibitem{CK06}V.~A.~Karmanov, J.~Carbonell, and M.~Mangin-Brinet,
Nucl.~Phys.~A {\bf 790} (2007) 598c, arXiv:hep-th/0610158.
\bibitem{CK07}V.~A.~Karmanov, J.~Carbonell, and M.~Mangin-Brinet,
Few Body Syst.~{\bf 44} (2008) 283, arXiv:0712.0971 [hep-ph].
\bibitem{CK08}J.~Carbonell, V.~A.~Karmanov, and M.~Mangin-Brinet,
Eur.~Phys.~J.~A {\bf 39} (2009) 53, arXiv:0809.3678 [hep-ph].
% Fermionic bound-state constituents:
\bibitem{CK10a}J.~Carbonell and V.~A.~Karmanov, PoS LC2010 (2010)
014, arXiv:1009.4522 [hep-ph].
\bibitem{CK10b}J.~Carbonell and V.~A.~Karmanov, Eur.~Phys.~J.~A {\bf
46} (2010) 387, arXiv:1010.4640 [hep-ph].
\bibitem{CK10c}J.~Carbonell and V.~A.~Karmanov, Few Body Syst.~{\bf
49} (2011) 205, arXiv:1012.0246 [hep-ph].

% Nakanishi representation of BS amplitude
\bibitem{Nakanishi}N.~Nakanishi, Phys.~Rev.~{\bf 130} (1963) 1230.

% Reduction to Schrödinger equation
\bibitem{Lucha91}W.~Lucha, F.~F.~Sch\"oberl, and D.~Gromes,
Phys.~Rep.~{\bf 200} (1991) 127.
\bibitem{Lucha:Oberwoelz}W.~Lucha and F.~F.~Sch\"oberl,
Int.~J.~Mod.~Phys.~A {\bf 14} (1999) 2309, arXiv:hep-ph/9812368.
\bibitem{Lucha:Dubrovnik}W.~Lucha and F.~F.~Sch\"oberl, Fizika B
{\bf 8} (1999) 193, arXiv:hep-ph/9812526.

% Instantaneous BS equation
\bibitem{Lucha05:IBSEWEP}W.~Lucha and F.~F.~Sch\"oberl,
J.~Phys.~G: Nucl.~Part.~Phys.~{\bf 31} (2005) 1133,
arXiv:hep-th/0507281.

% Salpeter equation
\bibitem{SE}E.~E.~Salpeter, Phys.~Rev.~{\bf 87} (1952) 328.

% Reduced Salpeter equation
\bibitem{Henriques76}A.~B.~Henriques, B.~H.~Kellett, and
R.~G.~Moorhouse, Phys.~Lett.~B {\bf 64} (1976) 85.
\bibitem{Jacobs87}S.~Jacobs, M.~G.~Olsson, and C.~J.~Suchyta III,
Phys.~Rev.~D {\bf 35} (1987) 2448.
\bibitem{Gara89}A.~Gara, B.~Durand, L.~Durand, and L.~J.~Nickisch,
Phys.~Rev.~D {\bf 40} (1989) 843.
\bibitem{Gara90}A.~Gara, B.~Durand, and L.~Durand, Phys.~Rev.~D
{\bf 42} (1990) 1651; {\em ibid.} {\bf 43} (1991) 2447 (erratum).
\bibitem{Lucha92C}W.~Lucha, H.~Rupprecht, and F.~F.~Sch\"oberl,
Phys.~Rev.~D {\bf 45} (1992) 385.

% Spinless Salpeter equation
\bibitem{Lucha92}W.~Lucha and F.~F.~Sch\"oberl,
Int.~J.~Mod.~Phys.~A {\bf 7} (1992) 6431.
\bibitem{Lucha94:Como}W.~Lucha and F.~F.~Sch\"oberl, in:
Proc.~Int.~Conf.~on {\em Quark Confinement and the Hadron
Spectrum}, edited by N.~Brambilla and G.~M.~Prosperi (World
Scientific, River Edge, New Jersey, 1995) p.~100,
arXiv:hep-ph/9410221.
\bibitem{Lucha04:TWR}W.~Lucha and F.~F.~Sch\"oberl, Recent
Res.~Devel.~Physics {\bf 5} (2004) 1423, arXiv:hep-ph/0408184.

% Quasipotential
\bibitem{Todorov}I.~T.~Todorov, Phys.~Rev.~D {\bf 3}, 2351 (1971).

% Geometric spectral inversion
\bibitem{inv1}R.~L.~Hall, Phys. Rev. A {\bf 50}, 2876 (1994). % flat bottoms
\bibitem{inv2}R.~L.~Hall, J. Phys. A: Math. Gen. {\bf 28}, 1771 (1995). % GSI
\bibitem{inv3}R.~L.~Hall, Phys. Rev. A {\bf 51}, 1787 (1995). % WKB
\bibitem{inv4}R.~L.~Hall, J. Math. Phys. {\bf 40}, 699 (1999). % constructive inv
\bibitem{inv5}R.~L.~Hall, J. Math. Phys. {\bf 40}, 2254 (1999). % inv ineq

% Geometric spectral inversion for singular potentials
\bibitem{FIS}R.~L.~Hall and W.~Lucha, J. Math. Phys. {\bf 52},
112102 (2011), arXiv:1111.1159 [math-ph].

\bibitem{GS}S.~J.~Gustafson and I.~M.~Sigal, {\it Mathematical Concepts of
Quantum Mechanics} (Springer, New York, 2006). [The operator
inequality is proved for dimensions $d\ge 3$ on p.~32.]
\bibitem{RS2}M.~Reed and B.~Simon, {\it Methods of Modern Mathematical Physics II:
Fourier Analysis, Self-Adjointness} (Academic Press, New York,
1975). [The operator inequality is proved on p.~169.]
\bibitem{flug}S. Fl\"ugge, {\it Practical Quantum Mechanics}
(Springer, New York, 1974). [The Hulth\'en potential is discussed
on p.~175.]
\bibitem{RS4}M.~Reed and B.~Simon, {\it Methods of Modern Mathematical Physics IV:
Analysis of Operators} (Academic Press, New York,
1975).
\bibitem{GF}I.~M.~Gelfand and S.~V.~Fomin, {\it Calculus of
Variations} (Prentice-Hall, Englewood Cliffs, 1963. [Legendre
transformations~are discussed on p.~72.]
\bibitem{Jensen}W.~Feller, {\em An Introduction to Probability
Theory and its Applications, Volume II\/} (John Wiley, New York,
1971). [Jensen's inequality is proved on p.~153.]
\end{thebibliography}
\end{document}